PAPER • OPEN ACCESS

# Photometric Investigation of Novae T Pyx, BT Mon and V574 Pup at Quiescence by using the 2.4-m Thai National Telescope

To cite this article: Ritthichai Thipboon et al 2017 J. Phys.: Conf. Ser. **901** 012003

View the article online for updates and enhancements.

## Related content

- A DETAILED PHOTOMETRIC AND SPECTROSCOPIC STUDY OF THE 2011 OUTBURST OF THE RECURRENT NOVA T Pyxidis FROM 0.8 TO 250 DAYS AFTER DISCOVERY
  F. Surina, R. A. Hounsell, M. F. Bode et al.

- Searching for Past Outbursts of Recurrent Novae
  Peter B. Robinson, Geoffrey C. Clayton and Bradley E. Schaefer

- IDENTIFYING AND QUANTIFYING RECURRENT NOVAE MASQUERADING AS CLASSICAL NOVAE
  Ashley Pagnotta and Bradley E. Schaefer

# Photometric Investigation of Novae T Pyx, BT Mon and V574 Pup at Quiescence by using the 2.4-m Thai National Telescope

**Ritthichai Thipboon[1], Metichai Kaewrakmuk[1], Farung Surina[2,4,*]**

**and Nuanwan Sanguansak[3,4]**

[1] Faculty of Education, Chiang Rai Rajabhat University , 80 Moo 9, Bandu,Muang Chiang Rai 57100, Thailand
[2] Faculty of Science and Technology, Chiang Rai Rajabhat University, 80 Moo 9, Bandu,Muang Chiang Rai 57100, Thailand
[3] School of Physics, Institute of Science, Suranaree University of Technology, Nakhon Ratchasima 30000, Thailand
[4] National Astronomical Research Institute of Thailand Chiang Mai 50200, Thailand

*E-mail address: sc_farung@crru.ac.th

**Abstract.** Recurrent novae (RNe) are novae with multiple recorded outbursts powered by a thermonuclear runaway. The outburst occurs on the surface of the white dwarf which accompanies with a late type main-sequence or giant secondary star transferring material onto the white dwarf primary star. They resemble classical novae (CNe) outbursts but only RNe has more than one recorded outbursts. RNe play an important role as one of the suspected progenitor systems of Type Ia supernovae (SNe) which are used as primary distance indicators in cosmology. Thus, it is important to investigate the outburst type of CNe and RNe and finally ascertain the population of objects that might ultimately be candidates for Type Ia SNe explosions. The proposal that RNe occupy a region separated from CNe in an outburst amplitude versus speed class diagram was adopted. Since the low amplitude results from the existence of an evolved secondary and/or high mass transfer rate in the quiescent system, RNe candidates should accordingly have low amplitude. We selected 3 preliminary targets including T Pyx, BT Mon and V574 Pup. Their amplitudes are not that low but the lowest amplitude that can be observed with Thai National Telescope (TNT). We obtained their magnitudes at quiescence using ULTRASPEC camera on the 2.4-m TNT. The positions of three targets on optical and near-infrared color-magnitude diagrams suggest that all three should have main-sequence secondary stars. This is true for T Pyx, whose secondary star has been confirmed its spectroscopy to be a main-sequence star, but not yet confirmed for BT Mon and V574 Pup.

## 1. Introduction
Classical novae (CNe) are interacting binary systems whose outbursts are powered by a thermonuclear runaway in accreted material on the surface of a white dwarf (WD). The secondary stars in such systems fill their Roche lobe and material is transferred onto the WD primary star via an accretion disk [1].

Recurrent novae (RNe) show many similarities to CNe, but have had more than one recorded outburst [2]. RNe are suggested as one of the possible "progenitor systems" of Type Ia supernovae (SNe) via the single degenerated scenario especially the Roche Lobe overflow [3].

Currently, there are only 10 known Galactic RNe among the approximately 400 Galactic CNe [4]. They group themselves into three distinct classifications based on the evolutionary status of their secondary stars whether to be main-sequence stars (MS-Novae), sub-giant star (SG-Novae), or red giant branch stars (RG-Novae) [4]. The optical emission from all quiescent nova systems is a composition of the emission from three components: the WD, the accretion disk, and the secondary where the contribution from WD in the optical region is expected to be negligible. Meanwhile the contribution of the accretion disk depends on several factors including the accretion rate, disk size, system inclination, and observed wavelength, the contribution from the secondary star is much more straightforward and simply depends on the type of star (i.e. mass, age, and metallicity) and again of course the observed wavelength [4,5].

Thus one can distinguish the RG-Novae and SG-Novae from the MS-Novae population based on solely their quiescent optical and near-IR (NIR) properties. Therefore our aim is to estimate the evolutionary status of the secondaries in our target nova systems by comparing the results derived from the nova positions on the optical and NIR Colour Magnitude Diagrams (CMD).

### 2. Section of Targets

On the basis of the evolved secondaries and/or high mass transfer rate in the quiescent system, the novae must be brighter at quiescence than those whose secondaries are main-sequence stars and then have a lower outburst amplitude as a result [6,7] (See [8] for details in target selection). We selected 3 preliminary targets including T Pyx, BT Mon and V574 Pup based on their outburst amplitudes of 9.1, 7.6, and 10.2 magnitude, respectively. Their amplitudes are not that low but the lowest amplitude that can be observed with Thai National Telescope (TNT) on 13 December 2015 was permitted. Among these three targets, T Pyx and BT Mon have been examined by this method previously by [8] but there are none of this investigation applied to V574 Pup.

### 3. Observations and Data Reduction and Analysis

Quiescent photometric magnitudes of the three targets in u', g',z', r', i' filter system were obtained using ULTRASPEC camera on TNT in the night of 13 December 2015. The data reduction processes were carried out by using IRAF packages. The apparent magnitudes in u', g',z', r', i' system then were transformed to UBVRI Johnson system using transformation equations provided by [9] and then derive absolute magnitudes by adopting distances from literature. Table 1 shows the optical magnitudes obtained by TNT while the 2MASS-JHK magnitudes were derived from catalogues [10]. Errors in $M_V$ were propagation errors from the observational errors and transformation equations. For T Pyx and BT Mon, their *B-V* colours were adopted from [11] and [12], respectively. Since V574 Pup lacks of *B-V* in literature therefore we estimate it from 3 different extinction maps from [13], [14], and [15].

Table 1. Observed optical and NIR apparent magnitudes ($M_V$) of three target novae at quiescence.

| Nova | d (pc) | $M_V$ | B-V | J | J-H | $M_J$ |
|---|---|---|---|---|---|---|
| T Pyx | 4800 (±500)[16] | 3.6 (±0.2) | 0.11 (±0.03) [11] | 15.15 (±0.05) | 0.19 (±0.09) | 1.75 (±0.4) |
| BT Mon | 1700 (±300)[17] | 0.19 (±0.4) | 0.18 (±0.4) [12] | 14.4 (±0.4) | 0.44 (±0.05) | 3.2 (±0.41) |
| V574 Pup | 4600 (±300)[18] | 3.9 (±0.2) | 0.3(±0.9) | 12.7 (±0.1) | 0.0 (±0.1) | -0.6 (±0.2) |

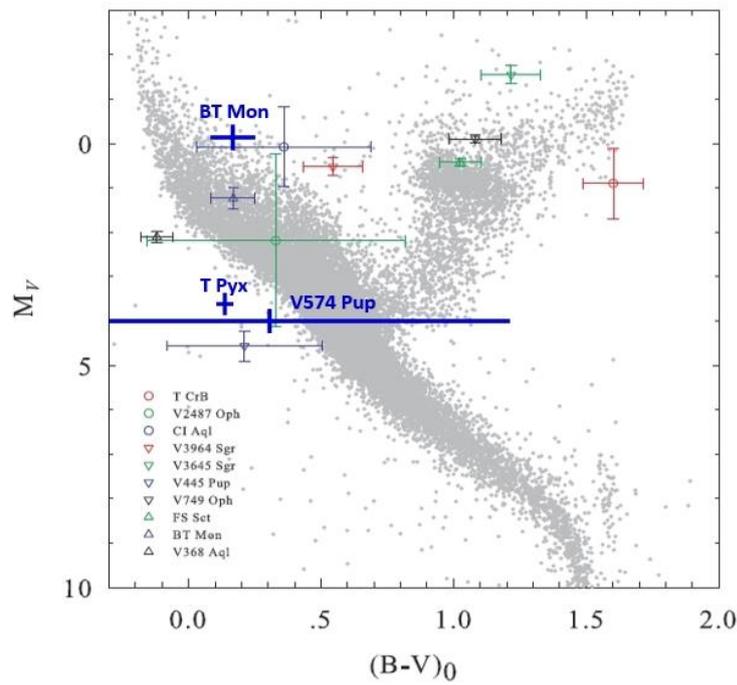

Figure 1: Optical CMD showing 3 targets (thick solid plus signs) compared with stars generated by cross-correlating the Hipparcos and 2MASS catalogues.

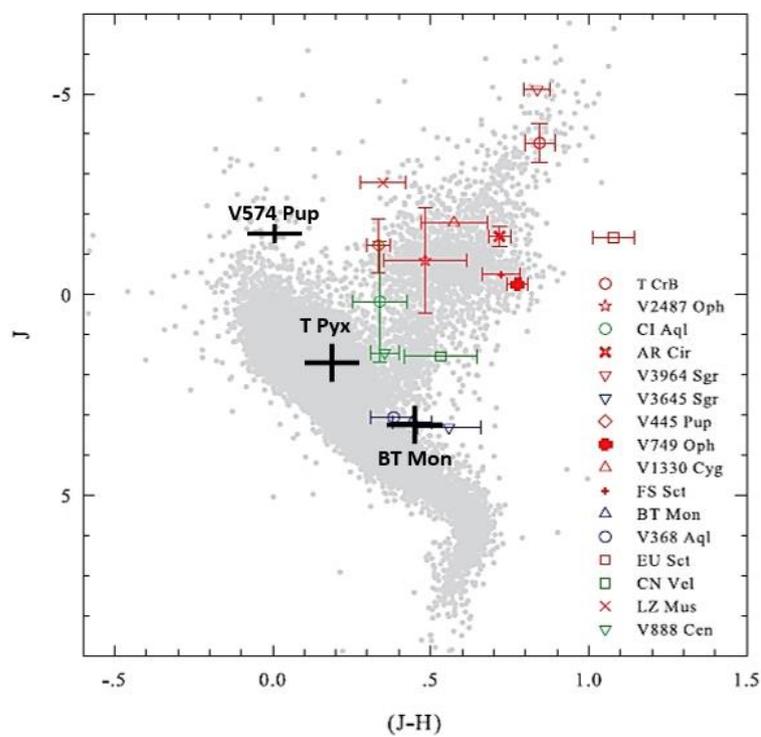

Figure 2: Near-Infrared CMD showing 3 targets (thick solid plus signs) compared with stars generated by cross-correlating the Hipparcos and 2MASS catalogues.

## 4. Results and Discussion

Comparing the positions of our novae on the visible and NIR CMDs to those presented in [4,12] where the RG-Novae, SG-Novae and MS-Novae are coloured in red, green and blue. Figure 1 shows that the optical CMD cannot distinguish among three classifications. While Figure 2. provides the clear separations among three types. The positions of three targets on the NIR CMD falling in the region of MS-Novae (shown in thick black plus signs) suggest that all three targets should have main-sequence secondary stars. This is true for T Pyx whose spectra has confirmed its evolutionary status of the secondary star. Spectra of BT Mon and V574 Pup still need to be investigated. Since at quiescence, the optical spectra of particular novae are suggested to be dominated by that of the red giant [4,7]. The absolute magnitude in V band of BT Mon derived by this proceeding is significantly brighter than the previous studies due to the difference of the adopted distance.  Together with the results from the previous study [8], this paper help us again to conclude that NIR CMD is more reliable than the optical and can be used to initially estimate the evolutionary status of secondary stars of novae together with the low outburst amplitude criteria.

## 5. References

[1] Starrfield, S., Iliadis, C., & Hix, W. 2008, in Classical Novae, ed. M.F. Bode & A. Evans (2nd ed Cambridge: Cambridge Univ. Press), 77
[2] Anupama, G. C. 2008, in RS Ophiuchi (2006) and the Recurrent Nova Phenomenon ASP Conference Series, ed. A. Evans et al. (San Francisco, ASP), 401, 31
[3] Li, W., et al. 2011, Nature, 480, 348
[4] Darnley, M. J., Ribeiro, V. A. R. M., Bode, M. F., Hounsell, R. A., and Williams, R. P, 2012, ApJ, 746, 61
[5] Surina, F., Bode, M. F., and Darnley, M. J., 2015, Publications of the Korean Astronomical Society, pISSN: 1225-1534 / eISSN: 2287-6936
[6] Duerbeck, H. W. 1988, A&A, 197, 148
[7] Anupama, G. C., & Mikolajewska, J. 1999, A&A, 344,
[8] Surina, F., Bode, M. F., and Darnley, M. J., 2013, The 11th Asian-Pacific Regional IAU Meeting 2011, NARIT Conference Series, Vol. 1, p161-164
[9] Lupton et al., 2005, AAS, 20713308L, available at http://classic.sdss.org/dr5/algorithms/sdssUBVRITransform.html
[10] Cutri, R. M., et al. 2003, The IRSA 2MASS All-Sky Point Source Catalog, NASA/IPAC Infrared Science Archive.
[11] Schaefer, B. (2010)
[12] Surina et al., 2017 (in prep.)
[13] Rowles, J., & Froebrich, D. 2009, MNRAS, 395, 1640
[14] Dobashi, K., Uehara, H., Kandori, R., Sakurai, T., Kaiden, M., Umemoto, T., & Sato, F. 2005, PASJ, 57, 1
[15] Schlegel, D. J., Finkbeiner, D. P., & Davis, M. 1998, ApJ, 500, 525
[16] Sokoloski, J. L., Crotts, Arlin P. S., Lawrence, Stephen, Uthas, Helena, 2013, ApJ, 770, 33
[17] Smith, D. A., Dhillon, V. S., Marsh, T. R., ASP Conference Series, 137, 477
[18] Burlak, M.A, 2008, Astronomical Letter, 34, 249

**Acknowledgments**
This work was funded by the National Astronomical Research Institute of Thailand (NARIT) and Chiang Rai Rajabhat Univeristy (CRRU). We thank professor Mike. F. Bode and Dr. Matt J. Darnley from Liverpool John Moores Univeristy for the great inspiration.